\documentclass[journal]{vgtc}                 

\ifpdf
  \pdfoutput=1\relax                   
  \usepackage{graphicx}                
  \DeclareGraphicsExtensions{.pdf,.png,.jpg,.jpeg} 
\else
  \usepackage{graphicx}                
  \DeclareGraphicsExtensions{.eps}     
\fi%

\graphicspath{{figures/}{pictures/}{images/}{./}} 

\usepackage{microtype}                 
\PassOptionsToPackage{warn}{textcomp}  
\usepackage{textcomp}                  
\usepackage{mathptmx}                  
\usepackage{times}                     
\usepackage{cite}                      
\usepackage{tabu}                      
\usepackage{booktabs}                  

\usepackage{bbm}
\usepackage{color,colortbl}
\usepackage{enumitem}
\usepackage{pifont}

\definecolor{pistachio}{rgb}{0.58, 0.77, 0.45}
\newcommand{\eg}{\emph{e.g.}}
\newcommand{\ie}{\emph{i.e.}}

\usepackage{caption} 
\usepackage{amsmath}
\usepackage{url}

\vgtccategory{Analytics \& Decisions}

\title{\emph{TimeTuner:} Diagnosing Time Representations for\\ Time-Series Forecasting with Counterfactual Explanations}

\author{Jianing Hao, Qing Shi, Yilin Ye, and Wei Zeng, \textit{Member, IEEE}}

\authorfooter{
  \item
    Jianing Hao, Qing Shi, Yilin Ye, and Wei Zeng are with the Hong Kong University of Science and Technology (Guangzhou), Guangzhou, China. Yilin Ye and Wei Zeng are also with the Hong Kong University of Science and Technology, Hong Kong SAR, China.
    E-mail: \{jhao768@connect., brantqshi@, yyebd@connect., weizeng@\}hkust-gz.edu.cn
  \item
    Wei Zeng is the corresponding author.
}

\abstract{
Deep learning (DL) approaches are being increasingly used for time-series forecasting, with many efforts devoted to designing complex DL models. 
Recent studies have shown that the DL success is often attributed to effective data representations, fostering the fields of feature engineering and representation learning.
However, automated approaches for feature learning are typically limited with respect to incorporating prior knowledge, identifying interactions among variables, and choosing evaluation metrics to ensure that the models are reliable.
To improve on these limitations, this paper contributes a novel visual analytics framework, namely \emph{TimeTuner}, designed to help analysts understand how model behaviors are associated with localized correlations, stationarity, and granularity of time-series representations.
The system mainly consists of the following two-stage technique: We first leverage counterfactual explanations to connect the relationships among time-series representations, multivariate features and model predictions. 
Next, we design multiple coordinated views including a partition-based correlation matrix and juxtaposed bivariate stripes, and provide a set of interactions that allow users to step into the transformation selection process, navigate through the feature space, and reason the model performance.
We instantiate \emph{TimeTuner} with two transformation methods of smoothing and sampling, and demonstrate its applicability on real-world time-series forecasting of univariate sunspots and multivariate air pollutants. 
Feedback from domain experts indicates that our system can help characterize time-series representations and guide the feature engineering processes.
}

\CCScatlist{
  \CCScatTwelve{Human-centered computing}{Visu\-al\-iza\-tion}{Visu\-al\-iza\-tion techniques}{Treemaps};
  \CCScatTwelve{Human-centered computing}{Visu\-al\-iza\-tion}{Visualization design and evaluation methods}{}
}


\begin{document}

\maketitle

\section{Introduction}
Time-series forecasting is a critical research problem with a wide range of applications, including finance, economics, and weather forecasting.
Accurate predictions of time-series data are vital in decision-making processes, allowing users to plan and allocate resources effectively.
However, traditional methods, such as autoregressive integrated moving average (ARIMA) and its variants, are limited in modeling nonlinear and complex time-series data.
Recent advances in deep learning (DL) have driven the development of more sophisticated time-series forecasting models, particularly long short-term memory (LSTM) networks.
LSTM has demonstrated remarkable performance in dealing with complex time-series data, as evidenced by its success in various applications such as predicting the spread of COVID-19~\cite{covid_2020}, analyzing financial data~\cite{petro_2019}, and forecasting economic trends~\cite{eco_2018}.

However, DL models are often criticized for being overly complex and lacking interpretability, and their behavior is highly sensitive to the representation of time-series data.
A recent study~\cite{dowe_2021} has shown that appropriate data processing techniques can improve the performance of time-series prediction models without relying on overly complex DL models.
In recognizing the significance of data representation, representation learning~\cite{representation_2013}, which focuses on developing automated approaches for learning better data representations, has become increasingly popular in the DL domain.
However, automatic approaches for representation learning alone have limitations in incorporating prior knowledge about intrinsic time-series patterns. For example, the sunspot dataset analyzed in the case study exhibits strong periodic patterns of approximately 13 months, which are not readily apparent in the feature space.
For multivariate time series, automated approaches alone are inadequate for uncovering interactions among variables~\cite{muhlbacher_2013_partition}.
There is a growing need for interactive solutions to increase user engagement in exploring the impact of different transformation methods on predictions, analyzing outliers, gaining more intuitive explanations, and subsequently selecting appropriate transformation methods.

Meeting this need is challenging due to two main reasons.
Firstly, current evaluation for time-series forecasting is typically limited to single numeric metric, such as root mean square error (RMSE) for the training/testing dataset as a whole, which overlooks the uniqueness of individual sliding windows and does not explicitly reveal which parts of the data representations are responsible for model behavior and resulting predictions~\cite{wexler_2019_what-if}.
Moreover, the lack of interpretability of DL models hinders users from understanding how different transformation methods affect predictions, preventing them from mining insights and understanding the reasons behind them.
Secondly, applying DL models to time-series forecasting requires considering a large number of possible data representations and vast volume of time-series data, even for a single representation.
The sliding window mechanism in data transformation also leads to a significant overlap in data.
Existing time-series visualization methods cannot address these issues adequately.

To overcome these challenges, this work contributes a visual analytics framework, namely \emph{TimeTuner}, leveraging counterfactual explanations to explain predictions for individual sliding windows, along with interactive time-series visualizations that are scalable to large-volume time series and multiple variables.
\emph{TimeTuner} comprises two-stage methods.
Firstly, in the \emph{forecasting and explaining} stage, raw time-series data are transformed into different representations using various transformation methods.
The representations are then passed to an LSTM model, and counterfactual explanation metrics, including SHapley Additive exPlanations (SHAP) and correlation, are derived.
Secondly, in the \emph{interactive exploration} stage, we develop a multiple-view visualization system that reveals the associations among variables, representations, and predictions from various perspectives.
Specifically, we propose a new design of juxtaposed bivariate stripes that can simultaneously depict bivariate representation and prediction metrics of multiple time-series representations.
We also leverage partition-based correlation matrices~\cite{muhlbacher_2013_partition}, horizon graphs~\cite{horizon_2009}, and other designs to address diverse analytical tasks.
To demonstrate the general applicability of \emph{TimeTuner}, we instantiate the system with two transformation methods of smoothing and sampling, and apply them to both univariate and multivariate time-series forecasting tasks on two real-world time-series datasets.
The case studies, along with feedback from domain experts, illustrate the effectiveness of our system.

In summary, our work makes the following contributions:
\begin{itemize}
    \item We develop \emph{TimeTuner}, a general visual analytics framework that integrates counterfactual explanations with interactive visualizations to diagnose the interactions among variables, representations, and predictions in time-series prediction.
    \emph{TimeTuner} is instantiated with two examples of common transformation methods to demonstrate the effectiveness of the framework.
    The source code of \emph{TimeTuner} is freely available at \url{https://github.com/CatherineHao/TimeTuner}.

    \item We propose a new design of juxtaposed bivariate stripes that provides a scalable visualization for comparing and exploring different time-series representations and corresponding predictions.
    
    \item We conduct two cases on univariate and multivariate datasets using \emph{TimeTuner}. Our analysis uncovers valuable insights, such as the sensitivity of predictions on abnormal events and the association between performance metrics and explanation metrics. 
    These insights provide useful hints for selecting appropriate transformation methods and explaining why models perform poorly.
\end{itemize}

\section{Related Work}\label{sec:related}

\textbf{Time-Series Forecasting}.
Traditional time-series forecasting methods focus on parametric models informed by domain knowledge, such as auto-regression (AR), ARIMA~\cite{ARIMA_1970}, and exponential smoothing~\cite{exp_1985}.
In recent years, DL has gained increasing popularity in time-series forecasting. 
Among various DL models, recurrent neural networks (RNNs) (\eg,\cite{deepstate_2018,rnn_2020}) have gained significant attention due to their capability to remember past information and model temporal dependencies.
However, when learning long-term dependencies, RNN architecture can suffer from gradient explosion and disappearance.
LSTM was developed to address these limitations~\cite{lstm_survey_2021}.
Nowadays, a plethora of works based on LSTM and its variants have been applied to time-series forecasting in various fields such as petroleum industry~\cite{petro_2019}, medical treatment~\cite{covid_2020}, and financial market~\cite{eco_2018}.

DL-based time-series forecasting is a complex process that requires careful consideration of data processing, feature engineering, and model development.
There has been a focus on developing advanced DL models, along with various visualization systems (\eg,\cite{strobelt_2018_lstmvis,kwon_2019_retainvis,ming_2020_protosteer,shen_2020_visual,mtseer_2021}) developed to facilitate sequence model understanding and improvement. 
However, challenges including limited improvement and generalizability issues remain unsolved~\cite{dowe_2021}.
Few efforts have been dedicated to data processing and feature engineering, which can also lead to improved predictions.
While automatic approaches such as representation learning~\cite{representation_2013} are popular, they have been criticized for their incompetence to incorporate prior knowledge and identify local interactions between variables~\cite{muhlbacher_2013_partition}.
Therefore, there is a need for an interactive visualization tool to analyze the impact of different transformation methods for time-series data, which is essential for accurate forecasting.
This work aims to fulfill this need by combining counterfactual explanations with visual analytics to meet diverse requirements.

\vspace{1mm}
\noindent
\textbf{Counterfactual Explanations}.
DL models are generally regarded as `black box' that is difficult to obtain human-comprehensible explanations directly from the models.
To address this challenge, a variety of explainable artificial intelligence (XAI) techniques have been proposed, which can be divided into two types: intrinsic interpretable models and post-hoc interpretation models~\cite{XAI_2018}.
Intrinsic interpretable models refer to models that have inherent interpretability, such as decision trees and linear regression.
However, the trade-off between interpretability and accuracy often hinders the usability of this class of methods~\cite{inter_2016}.
Post-hoc model explanation techniques are typically model-agnostic,
with some explaining a model's decisions by calculating feature attributions or through case-based reasoning,
and others providing counterfactual explanations to reveal what changes should have been made to the input instance to alter the outcome of an AI system~\cite{counter_2013,consur_2020, Mishra2022}.

Counterfactual explanations are hypothetical examples that explore the relationship between a model's input and output. 
By importing different inputs that could lead to different outputs, these hypothetical examples can help users understand how the model arrived at its predictions and understand the DL pipeline.
The \emph{What-If Tool}, developed by Google~\cite{wexler_2019_what-if}, supports analysis on both single data and the entire dataset, making it a versatile tool for exploring counterfactual explanations.
As counterfactual explanations do not require prior knowledge of deep learning, they are user-friendly to general users and have been applied in multi domains~\cite{dec_2018, con_2020,concl_2020}.
In the context of forecasting, counterfactual explanations can aid in determining which processing methods can be used to optimize future outputs (\eg,~\cite{zeng_2020}).
Combining them with interaction and visualization designs can help explore and interpret counterfactual explanations~\cite{mtseer_2021}.
For example, DCDE~\cite{DECE_2020} provides a set of interactions that enables users to customize the generation of counterfactual explanations to gain further insights.

In line with this research direction, our work employs counterfactual explanations for time-series forecasting, along with interactive visualizations that enable users to explore these explanations interactively.
This is nevertheless a challenging task because DL-based time-series forecasting involves components in various aspects, such as multivariate variables, diverse transformation methods, and large-volume time-series data.
A comprehensive exploration would require scalable time-series visualizations that can handle these tasks.

\vspace{1mm}
\noindent
\textbf{Time-Series Visualization}.
Appropriate visualizations are required to explore and analyze multivariate and large-volume time-series data~\cite{aigner_2011}. 
Line graph is a common technique for visualizing time series, allowing users to detect data trends and other patterns~\cite{line_2017}.
However, line graph can easily become unwieldy for complex time-series data.
Variations on line graphs, such as horizon graphs~\cite{horizen_2005}, stacked graph~\cite{stacked_2008}, and braided graphs~\cite{graphical_2010} have been developed to address the scalability issue.
These visualizations utilize vertical position or area to encode data values, which however, are constrained to the limited display space and human perception of minimal areas.
Other approaches are more scalable by mapping time-series data to pixel-wise visual elements and using color to encode their values, such as heatmaps~\cite{heat_2007,heat_2009}, sparkboxes~\cite{TL_2016}, and heat stripes~\cite{eye_2019}.
Calendar-based visualization~\cite{calendar_1999, calendar_2003} are also popular for visualizing periodic time-series data.

In addition to visualization techniques, interaction methods can be employed to address the scalability problem in exploring large-volume time-series data.
Overview+detail and focus+context are common interaction methods that help users examine time-series data with rich details while maintaining context~\cite{line_2006,ecglens_2018,lens_2010,newts_2023}.
Zooming is also a widely used technique, especially for exploring large time scales.
For example, TimeZoom~\cite{timezoom_2006} enables zooming to explore the arbitrary granularity of time units, while PlanningLines~\cite{planninglines_2005} allows zooming and panning for examining fine-grained plans with large time scales.

Our work aims to investigate the impact of various transformation methods on predictions through counterfactual explanations and an intuitive visual interface.
Given the extensive range of transformation methods and the potential overlap issues that arise from the sliding window mechanism, developing effective visualization techniques poses a significant challenge.
To overcome this obstacle, we propose a novel design of juxtaposed bivariate stripes, which is a modified version of heat stripes that offers the benefits of visualizing large-volume time series while also presenting bivariate representation and prediction metrics simultaneously using value-suppressing uncertainty palette (VSUP)\cite{VSUP_2018}.
Brushing and sorting supported in this design enable easy comparison and exploration of various representations and predictions.

\section{Background, Tasks, and System Overview}\label{sec:overview}

This section introduces the research background (Sect.~\ref{sec:background}), followed by analytical tasks (Sect.~\ref{ssec:anal_tasks}), and system overview (Sect.~\ref{ssec:system_overview}).

\subsection{Background}\label{sec:background}
Time-series forecasting is a fundamental problem in machine learning, particularly with the growing utilization of deep learning approaches.
These tasks can be categorized as single-step or multi-step, depending on the prediction horizon, and as univariate or multivariate, based on the number of attributes.
Since single-step forecasting is less common in practical applications, this work primarily focuses on multi-step time-series forecasting for both univariate and multivariate data.

Mathematically, this work considers a time-series dataset with $T$ time steps and $k$ variables, denoted as $\{X_t | t \in \{1, \cdots, T\} \}$, where each $X_t \in \mathbbm{R}^k$ denotes a set of observations at time step $t$.
For univariate time-series forecasting, $k=1$, while for multivariate time-series forecasting, $k>1$.
Our goal is to predict the values of the target variable, denoted by $y$, for time steps $T+1$ to $T+\triangle T$, based on the values of all $k$ variables at time steps $1$ to $T$.
Here, $y$ is the same variable with the input time series for univariate forecasting or one of the variables in the input time series for multivariate forecasting.
$\triangle T$ is the fixed forecasting length that varies for different tasks.
We can represent this as follows: 
$
\{y_{T+i}\}_{i=1}^{\triangle T} = f(\{X_t\}_{t=1}^T),
$
where $f$ is the mapping function, which can be learned using various machine learning techniques, such as RNNs, LSTMs, or other deep learning models, depending on the task's complexity and the data's nature.

\subsection{Analytical Tasks}\label{ssec:anal_tasks}

Based on our prior experience in utilizing LSTMs and a comprehensive review of existing research on representation learning for time series, 
we have observed that model developers are becoming increasingly dissatisfied with using a single overall metric for model performance.
Instead, they favor emerging counterfactual explanation techniques that support ``what-if analysis'' for instance. 
We first surveyed existing design studies for XAI, especially those using counterfactual explanations~\cite{DECE_2020, fairsight_2019,wexler_2019_what-if,explainer_2019,shen_2020_visual}, which emphasize the relationship between a model’s input and output, \ie, time-series representations and predictions. 
Next, we referred to existing studies on time-series forecasting, and identified key requirements such as identifying interactions between variables~\cite{muhlbacher_2013_partition,xai_2022,IML_2020}, which is essential for representation learning~\cite{representation_2013}.

In this way, we summarize three-level analytical tasks: \emph{T1) variable-level, T2) representation-level}, and \emph{T3) prediction-level exploration}.
Variable-level exploration involves examining the details of variables and identifying associations among them.

\begin{enumerate}[leftmargin=*, label=\textbf{T1.\arabic*}]
\vspace{-1.5mm}
    \item \textbf{Examine the details of variables.} 
    For effective counterfactual explanation, users need to inspect the details of variables, such as outliers of a variable and how different transformation methods affect them.

\vspace{-1.5mm}
    \item \textbf{Identify associations among variables.} In multivariate time-series forecasting, it is challenging to isolate a single variable and explain predictions independently. Therefore, users must first comprehend the associations between input variables to understand how these variables jointly influence the predictions.
\end{enumerate}

Representation-level exploration is paramount, which involves comparing different representations and analyzing their associations with variables and predictions.

\begin{enumerate}[leftmargin=*, label=\textbf{T2.\arabic*}]
\vspace{-1.5mm}
    \item \textbf{Compare different representations.}
    It is necessary to provide users with an \emph{overview} of all the available representations to enable efficient comparison, identify problematic representations, and select appropriate ones.
    To facilitate a deeper understanding of the varying quality of different representations, it is desirable to allow users to zoom in to \emph{details} of each representation at the finest level, namely specific windows with salient values, which may be the root cause of unexpected model performance.

\vspace{-1.5mm}
    \item \textbf{Analyze the association between representations and variables.}
    Users need to examine the relationship between variables and representations, particularly in multivariate time-series forecasting.
    Understanding how each variable is correlated with, emphasized, or neglected by certain representations can help explain why certain representations lead to specific results.

\vspace{-1.5mm}
    \item \textbf{Analyze the association between representations and predictions.}
    One key task for users seeking to explain models in terms of representations is to understand the effects of different representations on predictions.
    This involves analyzing the correlation between representations and predictions, focusing on localized correlations that can provide counterfactual explanations.
\end{enumerate}

Prediction-level exploration aims to overview the overall patterns and detect outliers of predictions.

\begin{enumerate}[leftmargin=*, label=\textbf{T3.\arabic*}]
\vspace{-1.5mm}
    \item \textbf{Overview the overall patterns of predictions.}
    Overviewing the status of the predictions, including changes of the predictions over time, and corresponding evaluation metrics, is a pre-requisite for downstream tasks such as outlier detection.
    \item \textbf{Detect the prediction outlier.}
To diagnose model failures such as underfitting and overfitting problems, it is important to assist users in quickly discovering outliers in the predictions for the subsequent counterfactual explanation.
These outliers could reveal potential issues with the model or data and guide further analysis and refinement of the model.
\end{enumerate}

\begin{figure}[t]
    \centering
    \includegraphics[width=0.42\textwidth]{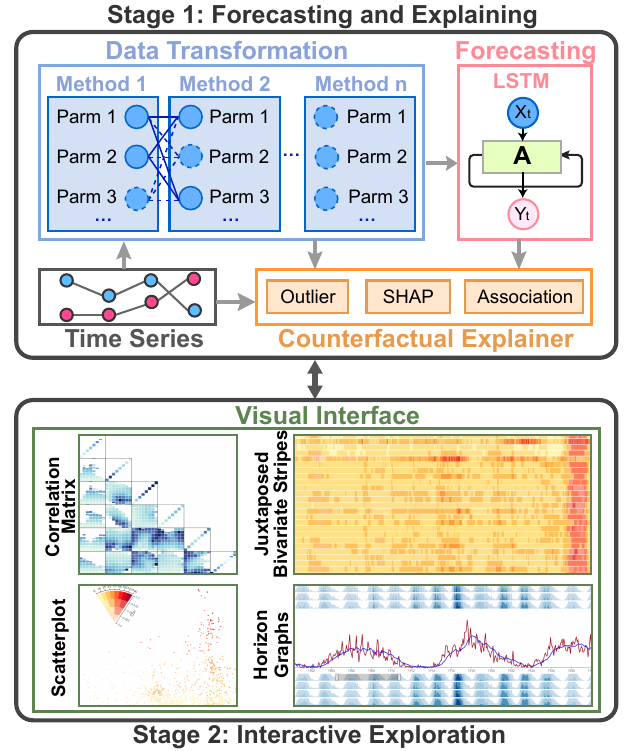}
    \vspace{-3mm}
    \caption{Overview of the \emph{TimeTuner} workflow, which mainly consists of four modules in two stages: 1) data transformation, 2) forecasting, and 3) counterfactual explaining in the \emph{forecasting and explaining} stage, and 4) visualization system in the \emph{interactive exploration} stage.}
    \vspace{-5mm}
    \label{fig:overview}
\end{figure}

\subsection{System Overview}\label{ssec:system_overview}
An overview of \emph{TimeTuner} is shown in Figure~\ref{fig:overview}, which mainly consists of four modules in two stages: 1) \emph{data transformation}, 2) \emph{forecasting} and 3) \emph{counterfactual explaining} in the \emph{Forecasting and Explaining} stage, and 4) \emph{visualization system} in the \emph{Interactive Exploration} stage.

In the first stage, raw time series are passed into the \emph{data transformation} module, yielding a variety of time-series representations.
The module can contain different transformation methods and optional parameters for time series. These methods affect properties of time-series representations, including localized correlations, stationarity, and granularity, and consequently affect the forecasting results.
Next, the representations are passed to the \emph{forecasting} module, where we choose the LSTM network as the forecasting model, a successful architecture ubiquitously applied to time-series forecasting.
Finally, the \emph{counterfactual explaining} module combines outlier detection, feature importance, and association analysis methods on raw and transformed time series, together with forecasting results, to support the counterfactual explanation of the forecasting model.

In the second stage, time-series representations, forecasting results, and explanation metrics are visualized in a coordinated multiple-view (CMV) system that mainly consists of the following: 
1) \emph{Variable Inspector View} presenting variable distribution and multi-variable correlations with a partition-based correlation matrix of Mosaic plots; 
2) \emph{Representation View}, displaying bivariate representation and prediction metrics for multiple representations with juxtaposed bivariate stripes color coded using VSUP; 
3) \emph{Prediction Comparator View}, depicting prediction distribution from the perspectives of explanation metrics using scatterplot; 
4) \emph{Temporal View}, visualizing time-series data details with horizon graphs.
A set of interactions are also provided, allowing users to explore the relationships among variables, representations, and predictions.
The visualization modules are implemented in D3.js~\cite{d3} and integrated using Vue.js~\cite{vuejs}, with Python Flask~\cite{flask} in the backend.

\section{Transformation, Forecasting, and Explanation}\label{sec:process}

\begin{figure}
    \centering
    \includegraphics[width=0.82\linewidth]{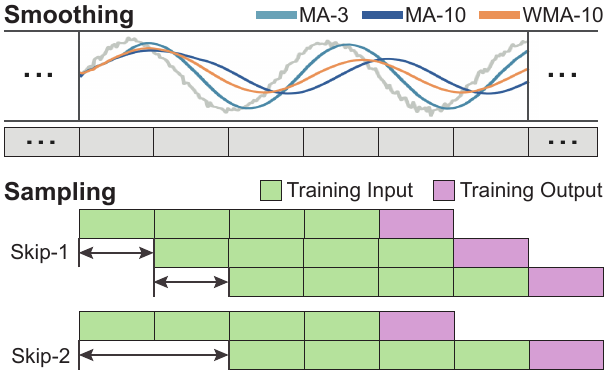}
    \caption{The data transformation includes two steps: 1) \emph{smoothing} using \emph{MA} and \emph{WMA} to remove noises and generate smooth time series, and 2) \emph{sampling} with different skip lengths.}
    \vspace{-5mm}
    \label{fig:skip}
\end{figure}

This section introduces methods of data transformation (Sect.~\ref{ssec:transformation}), models used for time-series forecasting (Sect.~\ref{ssec:modeling}), and metrics used for counterfactual explanation (Sect.~\ref{ssec:explainer}), utilized in this work.
\subsection{Data Transformation} \label{ssec:transformation}
Raw time-series data often contain deficiencies (\eg, outliers) that can significantly impact the accuracy of forecasting results.
It is necessary to transform the data and generate proper time-series representations before feeding them into a forecasting model.
In this work, \emph{TimeTuner} is instantiated with two transformation methods$-$smoothing and sampling, which are commonly used transformation methods for time-series data.
We aim to examine the impact of these two techniques on the forecasting results.
Specifically, we compare various commonly used smoothing and sampling methods as follows.

\begin{itemize}[leftmargin=*]
\item \textbf{Smoothing.} 
Smoothing is a widely-used technique to reduce or eliminate fluctuations in time series, which can increase the robustness of forecasting models against outliers.
This work evaluates two popular smoothing methods: \emph{Moving Average (MA)} and its variant \emph{Weighted Moving Average (WMA)}.
A simple moving average is calculated as the unweighted mean of the previous $m$ time-series data points: 
$
    MA_m = \frac{1}{m} \sum_{i=n-m+1}^{n} X_i,
$ where $m$ denotes the number of data points to be averaged. 
Instead of unweighted mean, \emph{WMA} assigns varying weights to the time-series data at different positions.
A typical example is to decrease weights in arithmetic progression when analyzing financial data.
The formula for \emph{WMA} is defined as follows:
$
    WMA_m = \frac{2}{m\times(m+1)} \sum_{i=n-m+1}^{n} X_i\times(n+1-i). 
$

Both \emph{MA} and \emph{WMA} use a smoothing parameter, denoted by $m$, to control the degree of smoothing.
A larger value of $m$ results in a smoother effect.
Users can choose different values of $m$ for different time series data and prediction tasks based on their prior knowledge.
    
\item \textbf{Sampling.}
After smoothing, a training dataset is needed for constructing DL-based forecasting models.
To make use of all available data, we employ the sliding window (SLICS) technique for sampling time series. The choice of an appropriate step size is crucial.
The step size determines the granularity of the sampled time-series data, and impacts the accuracy and precision of the forecasting results.
As shown in Figure~\ref{fig:skip} (bottom), a smaller step size generates more windows, providing a detailed picture of the time-series data. 
However, these windows may exhibit more variations that are difficult to learn.
Conversely, a larger step size may provide a more general view of time-series trends and patterns but may miss important changes or fluctuations between data points.
This work considers each sliding window as a time slice of the minimum analysis unit.
\end{itemize}

Each combination of smoothing and sampling methods yields a unique time-series representation, denoted as $\{X_t^s | t \in \{1, \cdots, \frac{T}{s}\}\}$,
where $X_t^s$ is the sliding window at position $t$, and $s$ is skip length utilized in the corresponding transformation method.

\subsection{Time-series Forecasting} \label{ssec:modeling}
Before applying the DL forecasting model, we perform the following tests on each time-series representation $X^s=\{X_t^s\}$:
\begin{itemize}[leftmargin=*]

\item
\textbf{Augmented Dicky-Fuller (ADF).}
The ADF test checks stationarity in time-series data, assuming that the time series follows a random walk model with drift~\cite{ADF_1995}.
Additional transformations are needed to achieve stationarity if the test fails.

\item
\textbf{Autocorrelation Function (ACF).}
The ACF value indicates the strength and nature of temporal dependencies between data points~\cite{ACF_1999}. 
The formula for calculating ACF at lag $m$ for a stationary time series $X^s$ is as follows:
\begin{equation}
    ACF(X^s,m) = \frac{\sum_{t=1}^n(X_t^s - \bar{X^s})(X_{t-m}^{s} - \bar{X^s})}{\sum_{t=1}^n(X_t^s - \bar{X^s})^2},
\end{equation}

where $\bar{X^s}$ denotes the mean value of $X^s$. 
\end{itemize}

\begin{figure}
    \centering
    \includegraphics[width=0.98\linewidth]{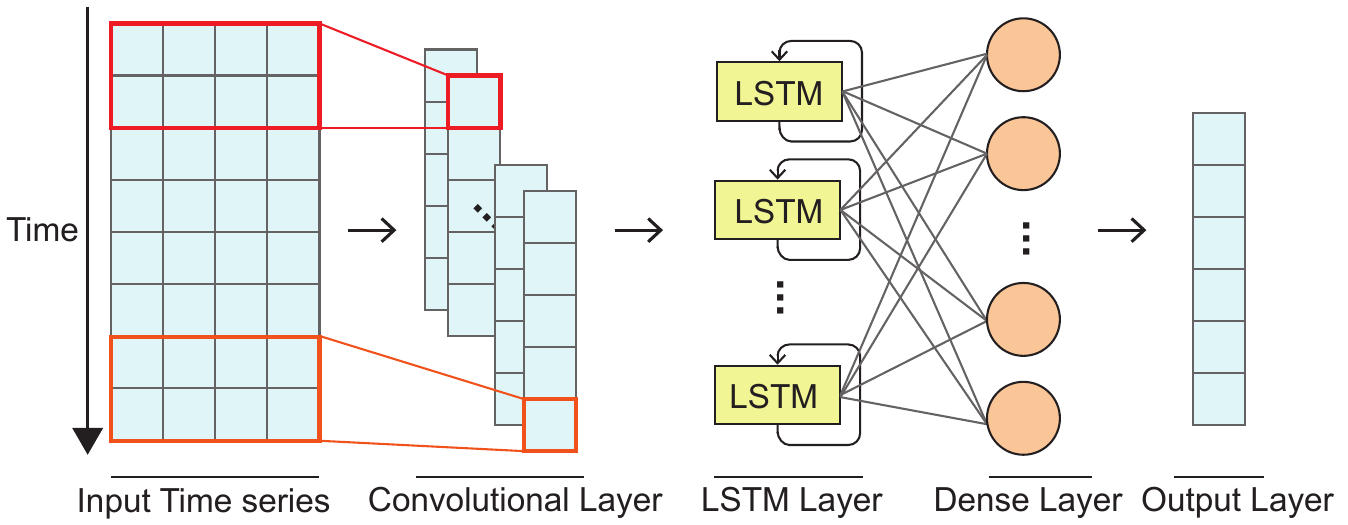}
    \caption{The simplified architecture of the LSTM model used for time-series forecasting in this work.}
    \vspace{-4mm}
    \label{fig:lstm}
\end{figure}

We adopt a simplified LSTM model as illustrated in Figure~\ref{fig:lstm}, implemented using the Keras framework~\cite{keras_link}.
The input matrix is a 3D array of shape (samples $\times$ time steps $\times$ features), where samples are the number of sliding windows; time steps correspond to the length of the training data within each window; features demote the number of input variables. 
We first apply the convolutional layer to detect and extract features from the input data.
The generated feature maps are then fed into the LSTM layer, which learns to identify important patterns and make predictions based on them.
Next, a dense layer and an output layer are used to make predictions using the learned features. 
To focus on the impact of different transformation methods rather than other experimental settings, we adopt consistent experimental settings for all transformations.
Specifically, we choose commonly used setups of 80:20 train-test data split ratio, and 100 epochs that reach converging state for the sunspots and air pollutants cases, to train a model for each time-series representation.
We choose a commonly used evaluation metric, \emph{RMSE}~\cite{RMSE_2005}, that can be applied to evaluate the performance of both a time-series representation and individual SLICS.

\subsection{Counterfactual Explanation}\label{ssec:explainer}
To provide actionable guidance to end-users in a user-friendly way.
\emph{TimeTuner} supports exploratory analysis of the impact of transformation methods on predictions by combining the counterfactual explanation metrics with a set of interactions to enable users to customize the generation of counterfactual explanation to find more insights and help their decision-making process.
In this section, we introduce the two metrics leveraged for the counterfactual explainer, including:

\begin{itemize}[leftmargin=*]

\item \textbf{SHapley Additive exPlanation value (SHAP)}~\cite{shap,shap_link} reflects the influence of the features in each sample.
Suppose the $i$th sample is $x_i$, the $j$th feature of the $i$th sample is $x_{i,j}$, the model's predicted value of the $i$th sample is $y_i$, and the baseline of the entire model is $y_{base}$ (usually the mean of the target variable for all samples), then the SHAP value $f(x_{i,j})$ of $x_{i,j}$ satisfies the following equation:
    \begin{equation}
        y_i = y_{base} + f(X_i,1) + f(X_i,2) + \cdots + f(X_i,k),
    \end{equation}
where $f(X_i,j)$ is the contribution value to $y_i$. 
$f(X_i,j)>0$ indicates that the feature $j$ has a positive effect on the predicted value, while $f(X_i,j)<0$ indicates negative effect.

\item \textbf{Pearson correlation coefficient (CORR.)}~\cite{corr_link} measures how strong the relationship between the predictions and sample is~\cite{CORR_2006}.
The Pearson correlation takes as input two data matrices X and Y:
    \begin{equation}
        CORR. = \frac{n \times \sum XY - \sum X \times \sum Y}{\sqrt{(n \times \sum X^2 - (\sum X)^2) \times(n \times \sum Y^2 - (\sum Y)^2))}},
    \end{equation}
    where $\sum XY$ is the sum of the products of samples and prediction values.
    CORR. ranges in $[-1,1]$, where 0 indicates no correlation and $+/-1$ indicates perfectly positive/negative correlations.

\end{itemize}

\section{Visualization Design}\label{sec:vis}

We develop a visual analytics system to address the analytical tasks (Sect.~\ref{ssec:anal_tasks}).
The system is expected to meet the following design criteria.

\begin{enumerate}[leftmargin=*, label=\textbf{DC\arabic*.}]
\item
\textbf{Coordinated Views.}
Various transformation methods generate representations distinct from each other, yielding diverse variable-, representation-, and prediction-level metrics.
To comprehensively reveal the associations between input representations and output predictions, coordinated multiple views (CMVs) that support visual analytics from multiple joint perspectives would meet the requirement.

\item
\textbf{Overview + Details.}
The visual analytics system should provide users with a clear overview of the input representations and output predictions while enabling detailed exploration as needed.
This can be achieved by combining aggregated visualizations, which provide a high-level overview, with unit visualizations that offer more granular details.

\item
\textbf{Flexible Interactions.}
Our system must offer rich interactions for flexible exploration.
Sorting time-series representations using various metrics can facilitate comparison while selecting and filtering operations can promote a more comprehensive \emph{overview + details} exploration. 

\end{enumerate}

Based on these design criteria, we design a CMV system that primarily incorporates five modules (Sects.~\ref{ssec:summary_view}$-$~\ref{ssec:sample_comp}), integrated with a set of interactions (Sect.~\ref{ssec:interaction}) to facilitate system exploration.

\subsection{Profile View}\label{ssec:summary_view}
\begin{figure*}[ht]
    \centering
     \includegraphics[width=.96\textwidth]{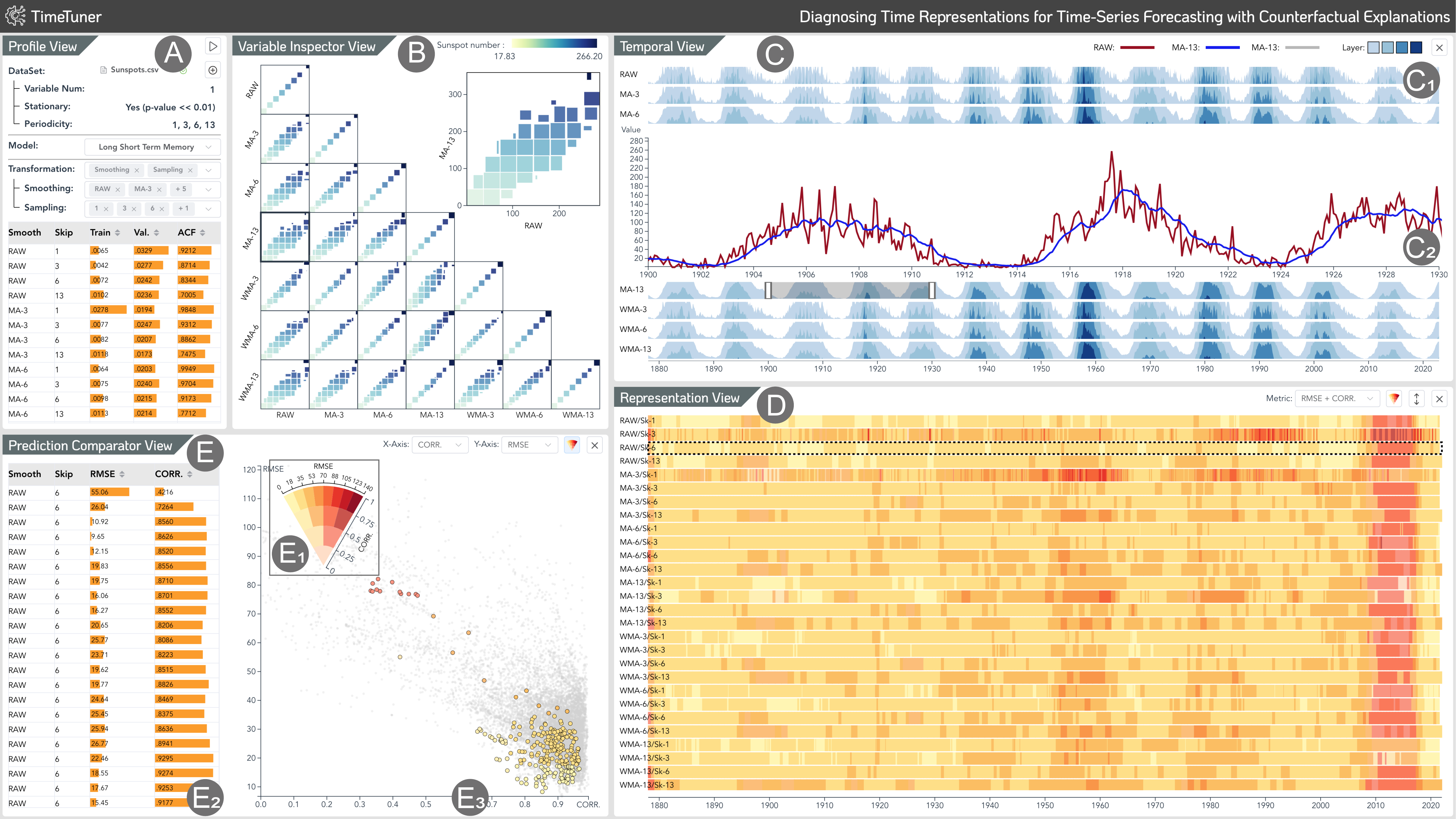}
    \caption{The interface of \emph{TimeTuner}. 
    \emph{Profile View} (A) provides information about the input dataset and an overview of all representations.
    \emph{Variable Inspector View} shows the localized distribution of different representations of univariate time series (B), helping users compare different smoothing methods.
    \emph{Temporal View} (C) presents global and local information of multiple representations/variables using horizon graphs.
    \emph{Representation View} (D) displays different time-series representations for users to explore, which shares the same color coding with \emph{Prediction Comparator View} (E).}
    \vspace{-3mm}
    \label{fig:interface}
\end{figure*}

The view allows users to upload a time-series dataset for exploration and presents an overview of available representations (\textit{T2.1}).
Statistics, including a number of variables, stationarity, and periodicity of the chosen dataset, are presented.
Users can also select one or multiple transformation methods and determine their parameters.
For example, in Figure~\ref{fig:interface} (A), the user selects two commonly adopted transformation methods of smoothing and sampling, and chooses time intervals of one month (1), one season (3), half a year (6), and meaningful periods (13) for sunspots advised by SILSO\footnote{http://www.sidc.be/silso/datafiles}.

The derived representations are presented in a table with sortable bars for users to have an overview of all representations and quickly identify representations of interest.
The table presents each representation as a row and shows its training error, validation error, and ACF value encoded by bar lengths.
Clicking on a row of interest will highlight the corresponding representation in \emph{Representation View} (Figure~\ref{fig:interface} (D)).
Users can sort the table according to different metrics.
In Figure~\ref{fig:interface} (A), the table shows that the representation `MA-3/Sk-1' (which stands for Moving Average with $m=3$ and skips with step size 1) has a large training error, indicating an abnormal representation for sunspot forecasting tasks.
Upon selection, the corresponding representation is highlighted in \emph{Representation View} (Figure~\ref{fig:interface} (D)).

\subsection{Variable Inspector View}\label{ssec:feature_inspector}
\begin{figure}[t]
    \centering
    \includegraphics[width=.98\linewidth]{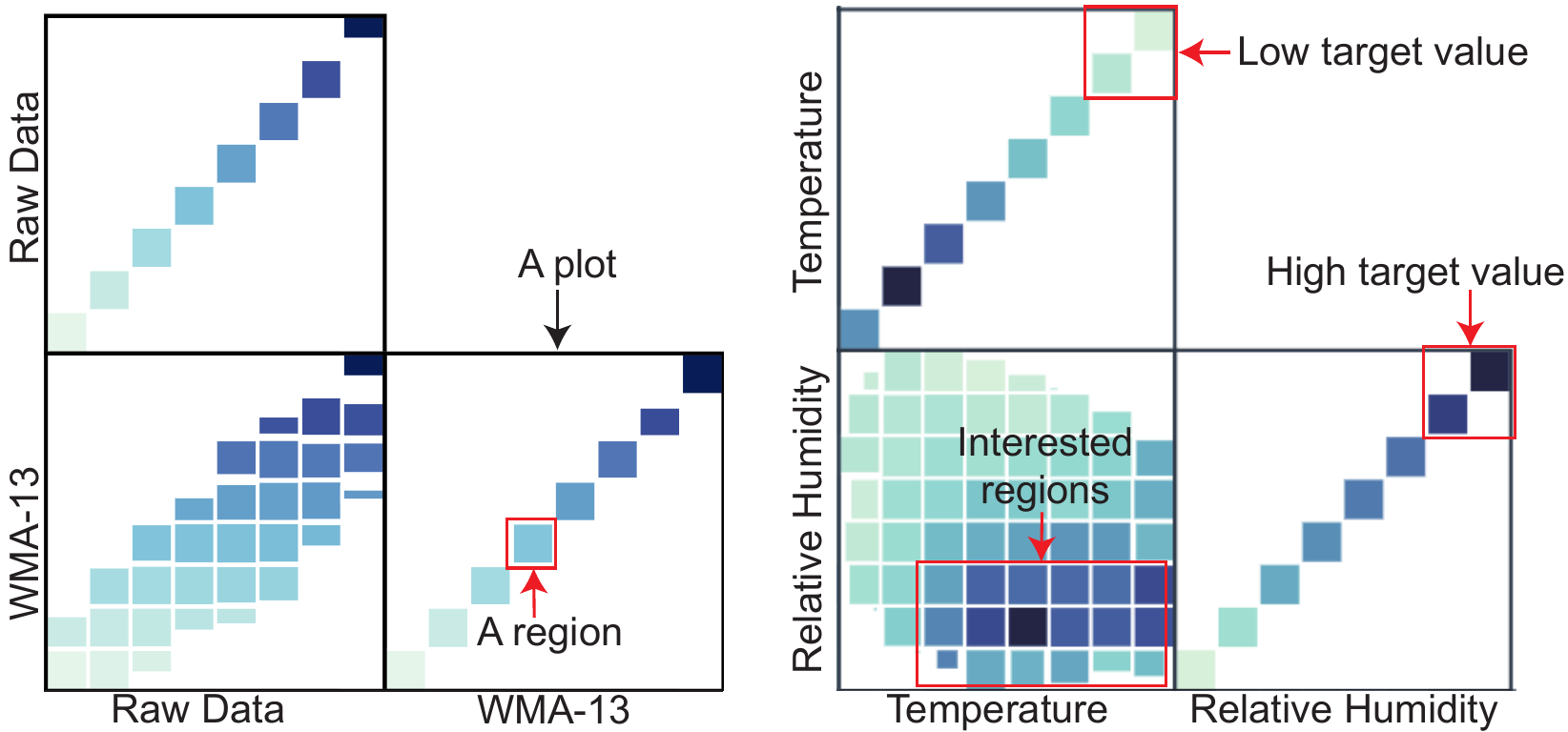}
    \caption{Left: correlation between `Raw' and `WMA-13' representations for univariate time-series data. Right: correlation between pairs of variables `Temperature' and `Relative Humidity' and the target variable `PM2.5' for multivariate time-series data.}
    \vspace{-8mm}
    \label{fig:feature}
\end{figure}

\emph{Variable Inspector View} (Figure~\ref{fig:interface} (B)) is designed to support the exploration of relationships between a feature space of continuous or discrete input variables and the target variable (\textit{T1.2, T2.2}).
Inspired by\cite{muhlbacher_2013_partition}, we provide a partition-based correlation matrix that partitions a variable into regions using uniform partitioning, and uses Mosaic plots to overview local relationships of pair-wise variables/representations.
The correlation matrix consists of multiple plots, each containing several rectangular regions.
The regions are colored based on the local relation metrics, with dark colors indicating a stronger correlation while no color for no correlation.
The view presents different information for univariate and multivariate time-series data as shown in Figure~\ref{fig:feature}. 

\begin{itemize}[leftmargin=*]
\item
For univariate data, this view shows the data distribution under different smoothing methods.
First, each time step in the time series $X_t$ is represented as a point in the scatterplot with its x \& y corresponding to the smoothed time series.
Then we aggregate the scatterplot into Mosaic plots which are uniformly partitioned into different regions, with the color of each region encoding the average values within the region.
When the data on X- and Y-axes are the same representation, only regions on the diagonal are colored.
The diverse distributions reflect the effects of different smoothing methods.
For instance, colors in the `Raw Data' and `WMA-13' plot (Figure~\ref{fig:feature} (left)) shift towards the upper right part, indicating that the time-series data smoothed by the `WMA' method would retain forward information making the data distribution shift.

\item
For multivariate data, the view visualizes correlations of all input variables that are ranked from top to bottom by variable importance. 
Within each plot, different input variables are represented on axes, while each region is color-coded by the value of the target variable.
By examining regions in each plot, users can mine the relationship between input variables and 
 the target variable, along with the relationship between pairs of variables and the target variable.
As shown in Figure~\ref{fig:feature} (right), the plots represent the relationship between `Temperature' and `Relative Humidity'.
The dark regions indicate high values of the target variable `PM2.5', while shallow regions indicate small `PM2.5' values.
The marked regions where `Temperature' is high and `Relative Humidity' is low indicate special patterns between `Temperature/Relative Humidity' pairs and PM2.5 values.
\end{itemize}

\subsection{Temporal View}\label{ssec:time_view}
\emph{Temporal View} (Figure~\ref{fig:interface} (C)) is designed to visualize the details of different time-series representations for univariate data (\emph{T2.1}) and variables for multivariate data (\emph{T1.1}).
It consists of two parts: horizon graphs for the context information and a line chart for the details.
\begin{itemize}[leftmargin=*]
    \item For the context information, each variable of the input time-series data is visualized as a 4-layer horizon graph~\cite{horizon_2009} to help users get a picture of data distribution.
    The horizon graph is a combination of area graph and histogram, allowing for more efficient use of screen real estate.
    As an example, it can be found that the number of sunspots dramatically increases in the 1960s (dark blue parts).

    \item For the details, we use a multiple-line chart with colors to encode different representations/variables selected by users.
    The selected time period is highlighted in the corresponding horizon graph.
    The line chart allows users to identify and analyze abnormal representations/variables by examining the details.
\end{itemize}

\subsection{Representation View}\label{ssec:representation_view}
\emph{Representation View} (Figure~\ref{fig:interface} (D)) is used to compare different representations (\textit{T2.1}) meanwhile depicting associations between representations and predictions simultaneously (\textit{T2.3}).
A key challenge here is to be scalable to too many time-series representations by different transformation methods and different parameters, each coupled with bivariate representation- and prediction-level metrics.
To address the challenge, we propose a novel design of juxtaposed bivariate stripes, with each time-series representation depicted as a strip and multiple representations juxtaposed from top to bottom.
All stripes share the same timeline with that in \emph{Temporal View}.
For instance, based on the skip length given in \emph{Profile View}, the view in Figure~\ref{fig:interface} (D) provides a total of 28 representations (seven smoothing alternatives $\times$ four sampling alternatives for the sunspot data).
For each slicing window in a representation, we compute its explanation metric (CORR. / SHAP value) as the first dimension and its performance metric (RMSE) as the second dimension.
The two-dimensional values of each window are encoded using a VSUP~\cite{VSUP_2018}.
As Figure~\ref{fig:interface} (E1) shows, VSUP is in a wedge shape.
We divide the values of the first dimension into eight ranges, while the values of the second dimension are divided into four ranges.
The use of VSUP encourages users to conduct more careful inspections of error-high slices and compare them.
We also support the exploration of changes in a single-dimension metric, by selecting the wanted metric through the drop-down box.
The VSUP palette will be changed to a sequential colormap correspondingly.

To enhance the view's scalability for large time-series data with numerous windows exceeding screen pixel capacity, we employ a two-stage optimization:
First, as discussed in Sect.~\ref{ssec:transformation}, because the skip length is always smaller than the span of a sliding window, the windows of a given representation have serious overlap on the time axis.
Therefore, if each window is represented by rectangles equal to its original length like in a normal heatmap, the colors of each sample will blend together and dilute the salient window.
To address the problem, we adjust the rectangles' length to fit the particular representation's skip length.
Each window is represented as a small rectangle aligned to the start of its corresponding sub-window.
This ensures that each skip has no intersection with others, effectively eliminating overlapping.

Second, we further employ a max-pooling aggregation method to ensure that metrics in each sliding window are covered.
Specifically, each pixel on the horizontal axis represents the aggregation of $n_{w}$ windows, where $n_{w}=\lfloor \frac{N_{w}}{P} \rfloor$, $N_{w}$ denotes the total number of windows and $P$ denotes the number of available pixels on the horizontal axis.
Then, in the max-pooling, the color encoding of the aggregated pixel is determined by the element-wise maximum of the bivariate metrics (\eg, RMSE vs. CORR.) for $n_w$ windows. 
The max-pooling aggregation can better preserve outliers than the commonly used average aggregation. 
We can identify some abnormal representations and predictions from the juxtaposed bivariate stripes.
As an example, the Raw/Sk-3 and MA-3/Sk-1 representations (Figure~\ref{fig:interface} (D)) are abnormal representations with large errors for all sliding windows, indicating that they are not suitable representations for this forecasting task.

\begin{figure}[t]
    \centering
        \includegraphics[width=.98\linewidth]{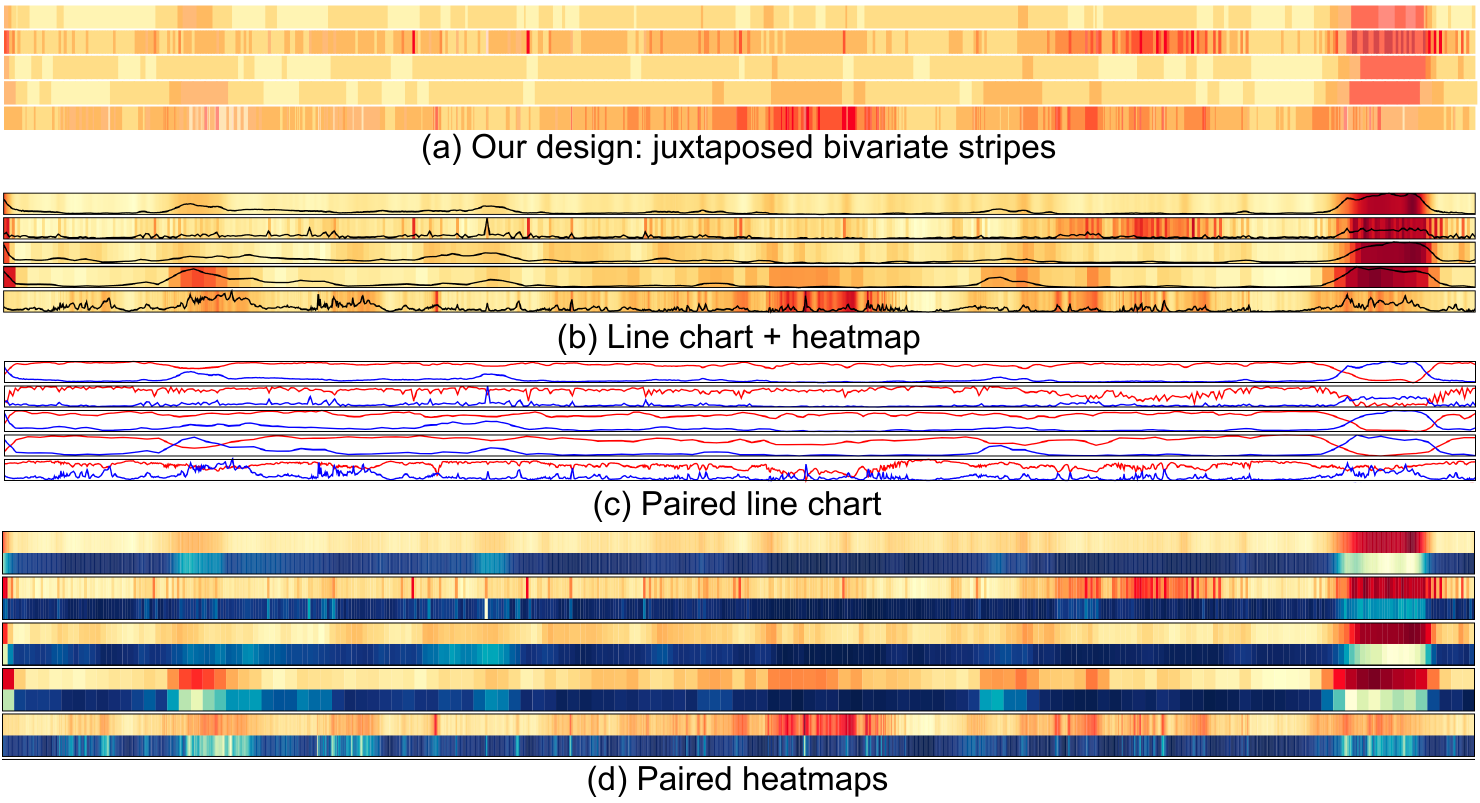}
        \vspace{-2mm}
    \caption{Our design of juxtaposed bivariate stripes (a) has better scalability than alternative designs: (b) line chart + heatmap, (c) paired line chart, and (d) paired heatmap.}
        \vspace{-7mm}
    \label{fig:rep_alt}
\end{figure}

\noindent
\textbf{Alternative Design}: As shown in Figure~\ref{fig:rep_alt}, we also considered three alternative design choices, including (b) line chart + heatmap, (c) paired line chart, and (d) paired heatmaps.
For all the designs, each juxtaposed row visualizes both the RMSE and CORR. of a specific representation.
Specifically, line chart + heatmap (Figure~\ref{fig:rep_alt} (b)) uses separate visual channels for different metrics by simply combining a line chart for one metric (\eg, CORR.) with color-encoded background (\eg, RMSE) for the other.
However, the background heatmap captures most of the attention due to the strong pre-attentive influence of colors, making it difficult to compare lines in different rows.
Paired line chart (Figure~\ref{fig:rep_alt} (c)) encodes RMSE and CORR. as two separate lines, making it difficult to compare and explore the metrics because their ranges are different.
Heatmaps are superior than line charts for displaying time-series data when the display space is limited~\cite{timeelide_2021}.
Another alternative design is paired heatmaps (Figure~\ref{fig:rep_alt} (d)) that arrange the bivariate metrics as two separate heatmaps, which would require two times more vertical space than our design using bivariate colormaps.

Therefore, these alternative designs are less scalable than the proposed bivariate juxtaposed stripes, especially when the number of representations increases due to the increasing number of transformation methods and their parameters.

\subsection{Prediction Comparator View}\label{ssec:sample_comp}
This view is designed to reveal the relationship between explanation metrics and performance metrics (\textit{T2.3}) and subsequently analyze prediction outliers (\textit{T3.1}).
To support the investigation of each prediction, we calculate its corresponding correlation coefficient, prediction error, and SHAP value.
In this way, each prediction can be represented as a point in the scatterplot, as shown in Figure~\ref{fig:interface} (E3).
Each point is encoded based on two-dimensional values schemed using the same VSUP with that in \emph{Representation View}.

Due to the large amount of predictions, displaying all of them would cause visual clutter.
Therefore, we use random sampling to reduce the overdrawing problem, along with lasso selection allowing users to see more local details by interaction.
The random sampling method can faithfully capture the density information of the whole distribution with the number of prediction points in each region proportional to local data density, which is important for the observation of distribution patterns and outliers. 
To see more details, users can select an interesting region using the lasso selection tool, highlighting all points in the region.
In addition, to show a more accurate prediction-level metric relationship, we provide a table (Figure~\ref{fig:interface} (E2)) adjacent to the scatterplot showing prediction errors (RMSE) along with correlation coefficients.
Users can sort the table according to different metrics for multi-faceted exploration.
The data filtering operation in the scatterplot and the table is coordinated, with lasso selection linked to row filtering of the table.

\subsection{User Interaction} \label{ssec:interaction}
In addition to the coordinated views, \emph{TimeTuner} also integrates various interactions to enable:
\begin{itemize}[leftmargin=*]
    \item \textbf{Selection \& Filtering}: 
    In \emph{Variable Inspector View}, users can hover over the interested plot to view more details in the upper right corner.
    If one plot is selected, the corresponding smoothed time series representation is displayed in the line chart of \emph{Temporal View}.
    Users can select a specific representation in \emph{Representation View}, predictions of interest in the table of \emph{Prediction Comparator View}, or filter a subset of predictions with the lasso tool in the scatterplot of \emph{Prediction Comparator View}.
    Selected/filtered prediction points are highlighted with black boundaries, while other points turn gray.
    In \emph{Representation View}, corresponding stripes are highlighted for users to explore abnormal predictions within the subset selected by the lasso tool.
    Users can select twice with the lasso tool in \emph{Prediction Comparator View} and get a union of highlighted prediction points.

    \item \textbf{Brushing}: Brushing is supported in \emph{Temporal View} for users to observe detailed changes in time series. When a particular data point is selected in the table of \emph{Prediction Comparator View}, the slider will be automatically moved to the corresponding time window.
    \item \textbf{Sorting}: 
    Users can sort representation or data points by certain metrics in the tables presented in \emph{Profile View} or \emph{Prediction Comparator View}, correspondingly.
    \emph{TimeTuner} also supports reordering the stripes in \emph{Representation View} by dragging certain stripes side-by-side to promote the comparison between the representations.

    \item \textbf{Scrolling}: \emph{Representation View} supports scrolling for displaying and comparing large representations. Since the view has limited space, displaying more than 30 representations can result in an excessively narrow height for each representation. When selecting multiple transformation methods and parameters, users can compare and explore the representations through a combination of scrolling and sorting.
\end{itemize}

\section{Evaluation}\label{sec:case}

\begin{figure*}[t]
    \centering
        \includegraphics[width=.98\textwidth]{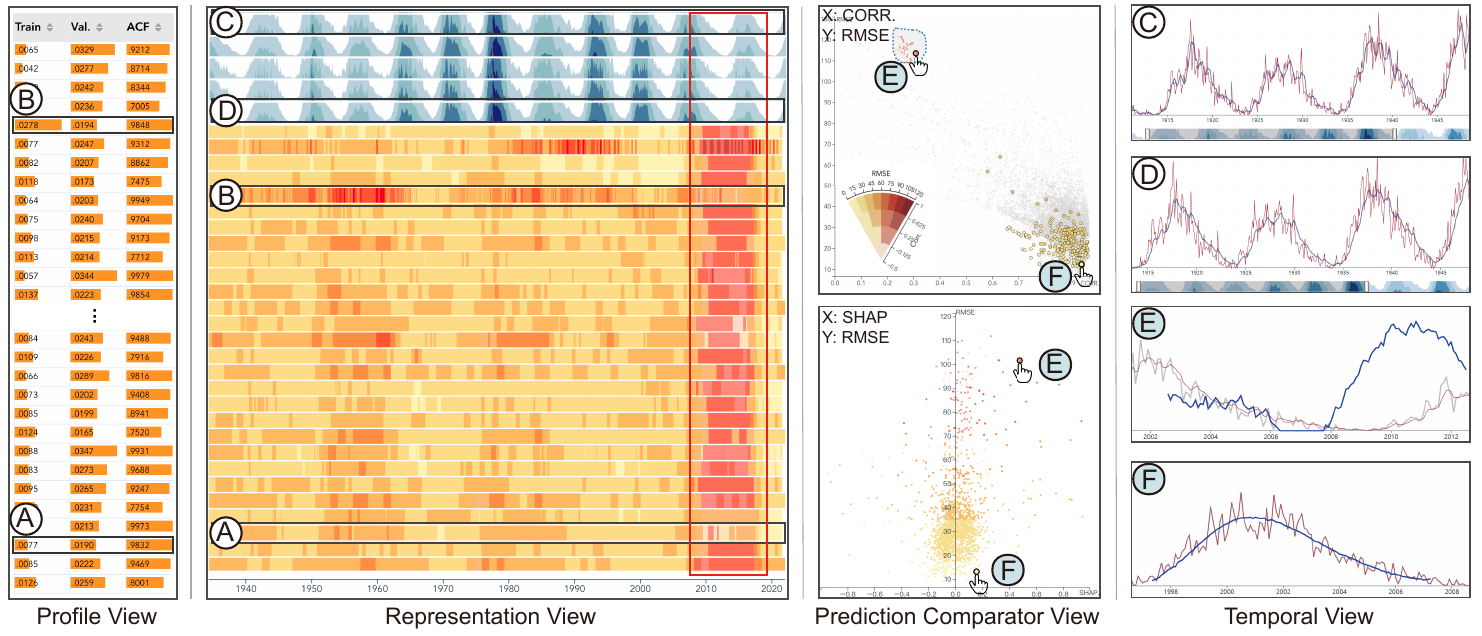}
    \caption{\emph{TimeTuner} can help identify abnormal representations and time slices in the case of forecasting sunspots. A, B, C, and D with the white tag are time-series representations, while E and F with the blue tag are focused time slices.}
    \vspace{-5mm}
    
    \label{fig:case1}
\end{figure*}

We conduct case studies $-$ one for univariate (Sect.~\ref{ssec:univariate_case}) and one for multivariate (Sect.~\ref{ssec:multivariate_forecast}) time-series forecasting task, and semi-structured expert interviews (Sect.~\ref{ssec:interview}) to evaluate the feasibility of \emph{TimeTuner}.
We illustrate these cases from the perspective of Mike, a system user, seeking a comprehensive understanding of the dataset and optimal transformation methods for the forecasting tasks.

\subsection{Case 1: Univariate sunspots forecasting}\label{ssec:univariate_case} 
In this case, Mike selects a famous univariate dataset ``Sunspot" that describes a solar phenomenon wherein a dark spot forms on the sun's surface.
The data is collected from the total sunspot number V2.0 published by SILSO.
The data include the monthly number of sunspots from January 1, 1818 to November 30, 2022.
The forecasting goal is to predict the number of sunspots in the next ten years (120 months) using sunspots from the past sixty years (720 months).

Firstly, Mike begins by selecting the smoothing methods and sampling parameters. He notices the periodicity of the sunspot dataset contains 1 month, 3 months, half a year, and 13 months, so he chooses the MA and WMA methods based on these four parameters and sets these four parameters as the skip length.
As Figure~\ref{fig:case1} shows, by sorting the prominent bar chart in Profile View, Mike can identify that the representation `WMA-13/Sk-3' achieves the highest prediction accuracy (Figure~\ref{fig:case1} (A)) and the representation `MA-3/Sk-1' has the largest training error (Figure~\ref{fig:case1} (B)).
Secondly, he wants to learn why the representation `WMA-13/Sk-3' is better suited than others.
So he clicks the bar and the corresponding representation is highlighted in \emph{Representation View}.
The representation `WMA-13/Sk-3' is the brightest one with significant SLICS errors only in the interval between 2009 to 2018 (red box).
He finds that almost all representations have large SLICS errors in this interval.
By observing the horizons in the \emph{Temporal View}, Mike realizes that this might be because 2009 is the period of minimum solar activity (the average number of sunspots in August 2009 was zero, the lowest recorded in 96 years since June 1913)~\cite{solar_2011}, and the model does not capture the weakening of sunspot activity.

Thirdly, Mike wants to select two predictions from different representations to inspect the counterfactual explanations.
He clicks on the horizon `MA-6' and `WMA-13' and then obtains the details of smoothed curves (Figure~\ref{fig:case1} (C) and (D)) in \emph{Temporal View}.
By observing the \emph{Prediction Comparator View}, he finds that there exists a negative relationship between CORR. and RMSE, with most of the predictions concentrated in the lower right corner.

For the predictions with obvious errors in the upper left corner (Figure~\ref{fig:case1} (E)), they correspond to the darkest SLICS in \emph{Representation View} (\textit{T2.3}).
By clicking on one prediction of the selected results, Mike focuses on the prediction values of this outlier window in \emph{Temporal View}, indicating that this presentation is unsuitable for this task.
Besides the CORR., Mike changes the X-axis to SHAP value and finds that predictions with small prediction errors are concentrated in the middle part, while the predictions with large prediction errors are concentrated on sides.
This situation is because each window's contribution to predictions is similar for models with good prediction results.
Significantly different contributions from some windows may indicate an overfitting problem.

\subsection{Case 2: Multivariate PM2.5 forecasting}\label{ssec:multivariate_forecast}
In this case, Mike selects a multivariate air pollutant dataset, which is collected by the Beijing US Embassy\footnote{https://aqicn.org/city/beijing/us-embassy/cn/}.
The data we used include the hourly data from 00:00 on January 1, 2013 to 23:00 on March 31, 2014 (15 months).
In this case, the forecasting task is to predict the values of PM2.5 in the next 12 hours using six variables, including PM2.5, temp (temperature), rh (relative humidity), psfc (SFC pressure), wnd\_dir (wind direction), wnd\_spd (wind speed) from the past 24 hours.
Outliers caused by specific events (such as Chinese New Year) are difficult to be captured and learned by models, resulting in poor performance of all models in the time range.

\begin{figure}[t]
    \centering
        \includegraphics[width=.96\linewidth]{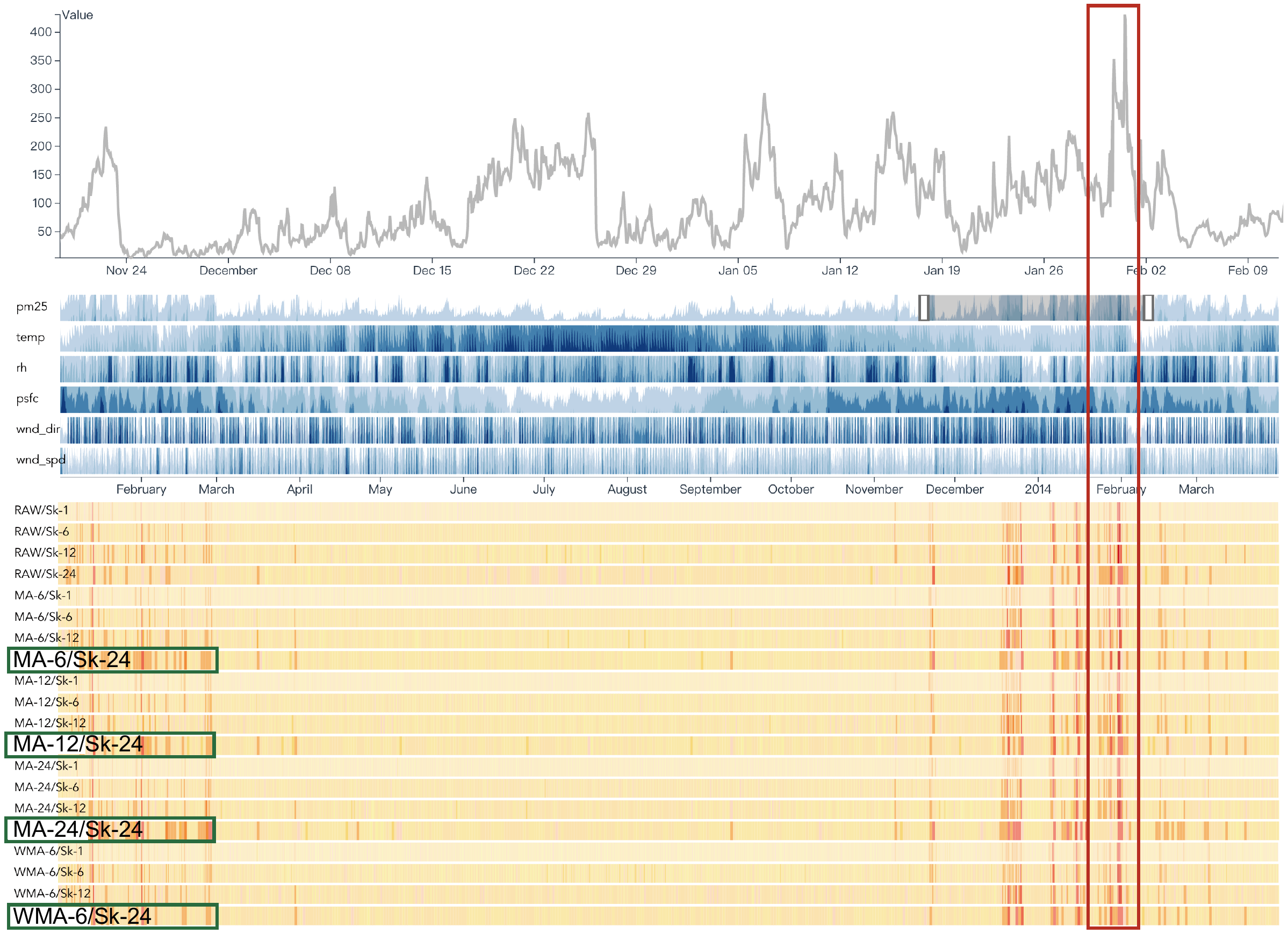}
    \caption{Compare the effects of different representations on predictions in forecasting PM2.5.} 
    \vspace{-6mm}
    \label{fig:case2-2}
\end{figure}

Firstly, Mike can easily find that the representation `WMA-24/Sk-1' is the most suitable one for the task by the sortable bar chart in \emph{Profile View} (\textit{T2.3}).
He then examines the \emph{Representation View} (Figure~\ref{fig:case2-2}) and identifies a time interval from January 30, 2014 to February 2, 2014 (red box) where all representations have high prediction errors, which he attributes to the setting off of fireworks during the CNY (\textit{T3.1}).
Secondly, Mike wants to analyze the differences among different sampling methods and determines that the representations of `Sk-24' have poor performances, as shown at the beginning of the timeline (green boxes).
After analyzing, Mike thinks it may be because the skip length is too large for these forecasting tasks, which ignores many typical patterns of the input dataset (\textit{T2.1}).

\begin{figure}[t]
    \centering
        \includegraphics[width=.94\linewidth]{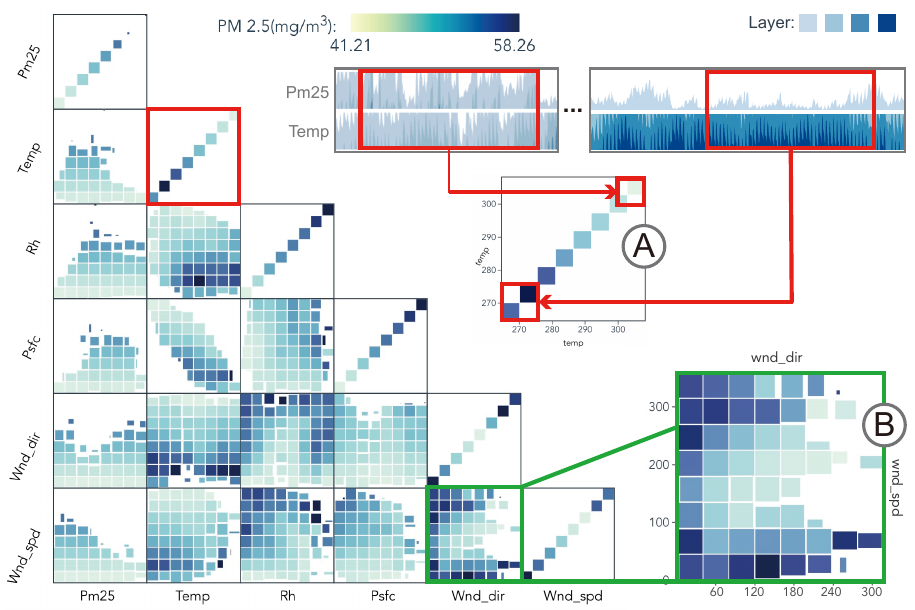}
    \vspace{-3mm}
    \caption{Identify the interactions among variables in the PM2.5 dataset.}
    \vspace{-8mm}
    \label{fig:case2-1}
\end{figure}

After exploring different representations, Mike wants to explore the relationship between different variable/variable pairs for PM2.5.
He begins exploration with \emph{Prediction Inspector View} (Figure~\ref{fig:case2-1}), and finds that there exists a negative correlation between PM2.5 and temperature (Figure~\ref{fig:case2-1} (A)), and a positive correlation with relative humidity.
Then he observes the horizon graph of PM2.5, temperature, and relative humidity and finds that the temperature is high in summer and the PM2.5 is low (red boxes), whilst PM2.5 is high in winter. 
Through inquiries, Mike learns that this may be due to two points: 1) \emph{Pollutant emissions}. 
In winter, coal is mostly used for heating in the northern region, and the amount of coal burned increases significantly.
What's more, due to the decrease in temperature in winter, the gas pressure and temperature of the working cycle of the automobile engine are not high, the combustion rate of the mixed gas slows down, causing incomplete combustion, and the exhaust emissions of motor vehicles increase, resulting in an increase in the emission of PM2.5 and its precursors (\eg, sulfur dioxide, nitrogen oxides, VOCs).
2) \emph{Temperature inversion}. At night in autumn and winter, the ground temperature drops sharply, so the temperature of the lower atmosphere close to the ground is very low, while the upper air does not cool down so quickly.
The temperature of the high layer is higher than that of the low layer, and the phenomenon of `temperature reversal' occurs. 
Once this cold and hot inversion layer is formed, the air cannot be convectioned up and down, making it difficult for pollutants to diffuse~\cite{pm25_2015}.

Besides the temperature and PM2.5 relations, Mike discovers intriguing insights in the `Wind direction - Wind speed' plot (Figure~\ref{fig:case2-1} (B)).
Low wind speeds hinder contaminant dispersion, resulting in high PM2.5 concentrations.
Surprisingly, even with high wind speeds, regions with south winds still exhibit elevated PM2.5 levels.
Mike speculates that the anomaly may stem from significant pollutant sources like factories and coal-fired power plants in the southern areas of Beijing.

\subsection{Expert Interview}\label{ssec:interview}
We conducted one-on-one interviews with five DL experts (\emph{DL1-DL5}) and five visualization experts (\emph{VE1-VE5}). All DL experts have experience in developing time-series forecasting models and can offer valuable feedback on visualization systems. The five visualization experts all have experience developing and working with interactive visualization systems and can provide valuable feedback on visual and interaction design from users' perspectives.
The interviews were held in a semi-structured format.
We first introduced the workflow of \emph{TimeTuner}. Then, we introduced the interface of \emph{TimeTuner} using the case examples from Sect.~\ref{ssec:univariate_case} and ~\ref{ssec:multivariate_forecast}. The introductory session took about thirty minutes.
Afterward, we asked them to explore \emph{TimeTuner} freely for fifteen minutes.
Finally, we collected their feedback.

\vspace{1mm}
\noindent
\textbf{Usability}
Experts (\emph{DL1}-\emph{DL5}) commended the combination of counterfactual explanations with visual analytics for analyzing the impact of time-series representations on LSTM-based predictions.
They acknowledged that time-series representations are as significant as deep learning models but have yet to receive more research attention.
All experts (\emph{DL1-DL5}) praised the system for its ability to support the exploration of both univariate and multivariate data.
According to a DL expert, ``\emph{the system helps me understand the input time-series data easily and select the appropriate time-series representations intuitively}".
Most experts agreed that it is reasonable to divide the overall goal into tasks at variable-, representation-, and prediction-level tasks, which \emph{TimeTuner} successfully accomplished.
However, \emph{DL2} had some confusion with the VSUP-based juxtaposed bivariate stripes, where she cannot identify sliding windows with smaller RMSE. 
After shown the stripes only for the RMSE metric in \emph{Representation View}, she appreciated that \emph{TimeTuner} provided unique insights and a complete exploration process for time-series forecasting tasks.

\vspace{1mm}
\noindent
\textbf{Visual and Interaction Design.}
\emph{TimeTuner's} visual and interaction design received positive feedback from both DL and visualization experts.
They found the views intuitive and well-aligned with data and analytical tasks.
The DL experts expressed particular enthusiasm for the horizon graphs and juxtaposed bivariate stripes, which were new to them. 
\emph{DL3} commented: ``\emph{These charts look impressive and have practical value in visualizing large-scale time series}."
Some visualization experts (\emph{VE1 \& VE3}) were unfamiliar with the VSUP palette~\cite{VSUP_2018} but appreciated its effectiveness over conventional bivariate colormaps in simultaneously presenting input features and prediction errors.
Three experts (\emph{DL2, DL3 \& DL5}) praised the selection and filtering operations, which helped them more effectively explore the relationship between representations and abnormal predictions.
Two experts (\emph{VE4 \& VE5}) found \emph{Variable Inspector View} slightly confusing for visualizing univariate time-series data.
After we clarified the meaning of each axis and color encoding, they were able to understand how to compare smoothing methods for univariate time series through this view.

\vspace{1mm}
\noindent

\vspace{-2mm}
\section{Discussion}
The case studies demonstrate how \emph{TimeTuner} aids users in comprehending the impact of minimal solar activity on sunspot forecasting (Sect.~\ref{ssec:univariate_case}), as well as the adverse effect of temperature on PM2.5 prediction (Sect.~\ref{ssec:multivariate_forecast}).
This is accomplished by providing counterfactual explanations for time-series representations and an interactive visualization system that enables users to investigate the connections among variables, representations, and predictions from different perspectives.
Domain experts verified the effectiveness of \emph{TimeTuner} in selecting appropriate representations for DL-based time-series forecasting.

\vspace{1mm}
\noindent
\textbf{Applicability}.
\emph{TimeTuner} is a versatile system that can be easily applied to other scenarios and DL models for time-series forecasting.
This is because \emph{TimeTuner} is model-agnostic and focuses on input representations and output predictions.
Specifically, we utilized smoothing and sampling methods commonly used for time-series transformation.
This process can be extended effortlessly to other transformation methods, including padding missed values, selecting features, etc., as suggested by a domain expert.
We used explanation metrics such as SHAP and CORR., which also apply to other tasks, including time-series classification and clustering.
Additional explanation metrics can also be integrated, such as partial dependence~\cite{friedman2001greedy} and summation-to-delta~\cite{shrikumar_2017_learning}.

\vspace{1mm}
\noindent
\textbf{Scalability}.
Currently, \emph{TimeTuner} relies on SHAP, which is time-consuming when calculating correlations among time slices, variables, and predictions.
As a result, all counterfactual explanation metrics are precomputed.
While some experts are willing to wait, allowing users to add new parameters and show the results interactively would be preferable.
This could be achieved using less computationally intensive metrics or implementing parallel computing.
From the visualization perspective, \emph{TimeTuner} is scalable to handle large-volume time series, benefiting from the innovative design of the juxtaposed bivariate stripes.
This design combines pixel-wise heat stripes with VSUP to achieve seamless encoding of bivariate values in times-series representations and predictions.
However, scalability issues can still arise when dealing with too many variables and representations of large-scale, large-volume time-series data.
When the user selects too many transformation methods and parameters in the control panel or the selected dataset has a large number of variables, the massive elements can cause the system to crash.
Our informal scalability testing suggests a minimum width of 20 pixels per plot/horizon/representation. It means that the variable number/smoothing methods should not exceed 9, and the number of representations should not exceed 30, otherwise may lead to visual disorder and difficult interactions.
To address this issue, we can use more robust preprocessing methods such as progressive clustering~\cite{calendar_1999} to combine similar time series and implement sophisticated visual and interaction designs, such as multi-foci navigation~\cite{zhao_2011_kronominer}.

\vspace{1mm}
\noindent
\textbf{Limitations and Future Work}.
\emph{TimeTuner} currently has several limitations that need to be addressed in future work.
First, our design of the \emph{Representation View} and \emph{Prediction Comparator View} may lead to confirmation bias because we use a backward-oriented reasoning process that acquires the diagnostic predictions first and then looks for supporting evidence.
To mitigate this issue, we need to investigate whether our visualization designs may alleviate potential risks and whether cognitive biases affect the experts' decisions. This can be done by conducting user studies to assess the precision of users' selection and cognition when using \emph{TimeTuner}.
Second, we offer limited time-step parameters in the sunspot and PM2.5 forecasting tasks, which may not be enough for some other scenarios.
To test the universality of \emph{TimeTuner} on more types of time-series data and models, we need to develop more computation-efficient metrics, robust preprocessing methods, and sophisticated visual and interaction designs to address scalability issues.
Furthermore, generating counterfactual explanations for various types of data and questions is an active research problem that we aim to support in the future.
We plan to open-source the code for time-series transformation and the visualization system to promote future research in this direction.
\vspace{-0.6mm}

\section{Conclusion}
This work highlighted three key challenges that limit the effectiveness of automated approaches for representation learning in time-series forecasting: the limitation of incorporating prior knowledge, identifying interactions among variables, and choosing evaluation metrics to ensure the model's reliability.
To overcome these limitations, we introduced a general visual analytics framework called \emph{TimeTuner}, which can help analysts understand model behaviors and associations at the variable, representation, and prediction levels.
We demonstrated the effectiveness of our approach in terms of in-depth insights learned in two case studies and positive feedback received from expert interviews.
Our findings suggest that visual association with localized correlations, stationarity, and different time-series representations significantly improve users' ability to interpret model performances and improve engineering processes. 
Overall, our work has highlighted the significant potential of counterfactual explanations combined with visual analytics in advancing the field of data representation learning.

\acknowledgments{
The authors wish to thank anonymous reviewers for their constructive comments. This work was supported in part by the National Natural Science Foundation
of China (No. 62172398) and Guangdong Basic and Applied Basic Research Foundation
(2021A1515011700).}

\bibliographystyle{abbrv-doi}
\bibliography{template}

\end{document}